# Uniaxial and hydrostatic pressure effects in $\alpha$-RuCl$_3$ single crystals via thermal-expansion measurements

Mingquan He,[1, *] Xiao Wang,[2, *] Liran Wang,[1] Frédéric Hardy,[1] Thomas Wolf,[1]
Peter Adelmann,[1] Thomas Brückel,[2, 3] Yixi Su,[2] and Christoph Meingast[1, †]

[1]*Institute for Solid State Physics, Karlsruhe Institute of Technology, 76021 Karlsruhe, Germany*
[2]*Jülich Centre for Neutron Science (JCNS) at Heinz Maier-Leibnitz Zentrum (MLZ),
Forschungszentrum Jülich GmbH, Lichtenbergstr. 1, D-85747 Garching, Germany*
[3]*Jülich Centre for Neutron Science (JCNS) and Peter Grünberg Institut (PGI),
Forschungszentrum Jülich GmbH, D-52425 Jülich, Germany*

(Dated: 12/22/17)

We present high-resolution thermal-expansion and specific-heat measurements of single crystalline $\alpha$-RuCl$_3$. An extremely hysteretic structural transition expanding over 100 K is observed by thermal-expansion along both crystallographic axes, which we attribute to a change of stacking sequence of the RuCl$_3$ layers. Three magnetic transitions are observed, which we link to the different stacking sequences. Using our data and thermodynamic relations, we derive the uniaxial and hydrostatic pressure derivatives of all three magnetic transitions. Our results demonstrate that magnetic order should be totally suppressed by very moderate pressures of 0.3 GPa to 0.9 GPa. Finally, we discuss why our results differ from recent hydrostatic pressure measurements and suggest a possible route to reaching the spin-liquid state in $\alpha$-RuCl$_3$.

## I. INTRODUCTION

The emergence of a quantum spin-liquid-state in a frustrated magnetic system has recently been theoretically solved by Alexei Kitaev in a honeycomb lattice model [1]. The possible realization of fractionalized Majorana-fermion excitations in such a system has initiated a large effort to experimentally search for such a physical realization of spin liquid in condense matter physics [2–5]. Experimentally, a few candidates, including iridates Na$_2$IrO$_3$ [6, 7], $\alpha, \beta, \gamma$-Li$_2$IrO$_3$ [8–12] and ruthenate $\alpha$-RuCl$_3$ [13–22], have recently been found. These materials all share a layered hexagonal lattice structure with weak interlayer coupling, which currently constitute the closest realization of Kitaev interactions [23]. However, all of the above compounds order magnetically at finite temperatures due to residual interactions. Pressure or doping, which are excellent tuning parameters in many systems, possibly could be used to force these near-Kitaev materials into a true spin-liquid ground state.

Among the Kitaev candidates, $\alpha$-RuCl$_3$ is of particular interest [2–4, 13]. In this layered material, edge sharing RuCl$_6$ octahedra form the layers in the $ab$ plane, that are stacked along the crystallographic $c$ direction via weak van-der-Waals bonds. The Ru$^{3+}$($4d^5$) ions have an effective spin-1/2 state, which orders into a zig-zag type antiferromagnetic(AF) ground state at finite temperatures [14, 16]. The magnetic transition temperature ranges from $\sim$ 7 K to $\sim$ 15 K depending on the stacking of the layers [14], which strongly affects the interlayer magnetic interactions. In fact, the low temperature crystal structure is still under debate; both trigonal $P3_112$ [18, 24], monoclinic $C2/m$ [14, 16, 21] and rhombohedral $R\bar{3}$ [18] types have been reported, suggesting that these structures are energetically nearly degenerate. Further, a structural phase transition with a large hysteresis below room temperature has been observed by magnetization [17], X-ray [18], and Raman scattering [15], the origin of which is however unclear.

In this article, we report on high-resolution thermal-expansion measurements of $\alpha$-RuCl$_3$ single crystals for the first time. A structural transition exhibiting a huge hysteresis is clearly observed along both crystallographic axes in the thermal expansion data, which we argue is due to a temperature induced change of stacking sequence. Further transitions are observed in both thermal expansion and heat capacity at $T_{N1} \sim 14$ K, $T_{N2} \sim 10$ K and $T_{N3} \sim 7$ K, which we assign to magnetic transitions occurring in the different polymorphs. Using thermodynamic relations, we predict that all three transitions will be suppressed by both uniaxial, as well as by hydrostatic pressure at similar rates. The resulting P-T phase diagram, which however differs significantly with direct hydrostatic pressure experiments [19, 25], suggests that it should be possible to stabilize the spin liquid state in $\alpha$-RuCl$_3$ using a very moderate pressure of roughly 0.3 GPa.

## II. METHODS

$\alpha$-RuCl$_3$ single crystals were grown by an evaporation/condensation technique in a temperature gradient. First, anhydrous RuCl$_3$ powder was sealed in an evacuated quartz glass ampoule. Then the ampoule was placed in a vertical tube furnace, keeping the starting powder at about 1000 °C. $\alpha$-RuCl$_3$ crystals grew at the colder end of the ampoule at about 960 °C within five days. The thermal-expansion of the $\alpha$-RuCl$_3$ single crystals was characterized by a home-built high resolution capacitance dilatometer [26]. The heat capacity measurements down to 400 mK were performed in a 14 T Quantum Design Physical Property Measurement System with a He-3 insert.

---

[*] These authors contributed equally to this work.
[†] christoph.meingast@kit.edu

## III. RESULTS

### A. Structural transition

The thermal-expansion, $\frac{\Delta L_i(T)}{L_i(300K)} = \frac{L_i(T)-L_i(300K)}{L_i(300K)}$, of a $\alpha$-RuCl$_3$ single crystal (Sample 1) along $a$ and $c$ axes are presented in Figure 1(a). The $c$-axis thermal-expansion is significantly larger than that of the $a$-axis, as expected for a weakly bonded layered material and in agreement with x-ray diffraction experiments[18]. Step-like features are clearly seen in the thermal-expansion along both directions, clearly indicating a structural transition. The transition is extremely hysteretic, and thus the transition temperature is not well defined. Upon cooling, the transition happens at $T_S^{cooling} = 50$ K (middle point of the jump) and $T_S^{cooling} = 66$ K for $a$-axis and $c$-axis, respectively. In contrast, the transition occurs at much higher temperature upon heating with $T_S^{heating} = 160$ K ($a$-axis) and $T_S^{heating} = 168$ K ($c$-axis). The slight difference in temperatures along different axes is most likely due to the uniaxial pressure applied by the cell, which is expected to increase (decrease) the transition temperatures along $c$-axis ($a$-axis), as explained in more detail below. The hysteric region spans more than 100 K, and similar hysteric behavior has also been captured by magnetization [17], X-ray [18], and Raman scattering [15] experiments. Such a large hysteresis is most likely due to a stacking rearrangement to a different crystallographic structure at low temperature (trigonal $P3_112$ or rhombohedral $R\bar{3}$ phase[18, 24, 27]). Unavoidable during such a transition is also a high degree of stacking faults and a likely fraction of non-transforming phase at low temperature (monoclinic $C2/m$[16, 21]). Thus it is not surprising that the low temperature structure is ill-defined and this probably explains the controversial low temperature crystal structures[16, 18, 21, 24, 27], as well as the multitude of magnetic transitions as described in the following section. However, no anomaly indicating a structural transition could be found in the heat capacity upon cooling as shown in Fig. 1(b), which demonstrates that the low- and high-temperature phases are energetically nearly degenerate.

### B. Magnetic transitions

Figure 2 shows the detailed thermal-expansion and specific heat data of two $\alpha$-RuCl$_3$ samples in the vicinity of the magnetic transitions. To better resolve the tran-sitions, the linear thermal-expansion coefficients, $\alpha_i = \frac{1}{L_i}\frac{dL_i}{dT}$, are shown in Figs. 2(a)(c). Anomalies in thermal-expansion coefficient $\alpha$ and specific heat $C$ represent signatures of phase transitions, and one can identify three transitions, which have previously been associated with antiferromagnetic transitions [14, 16, 17, 22]. In Sample 1, $T_{N1} = 13.5$ K, $T_{N2} = 9.8$ K and $T_{N3} = 6.7$ K [Figs. 2(a)(b) ], while $T_{N1} = 14$ K, $T_{N2} = 10.2$ K and $T_{N3} = 7.3$ K for Sample 2 [Figs. 2(c)(d)]. Both the transition temperatures and the strength of the anomalies vary from sample to sample, as has been observed previously [14, 16, 17, 22]. All transitions in Sample 2 appear at slightly higher temperatures and the anomaly at $T_{N3}$ of Sample 2 is significantly stronger than that of Sample 1 as seen both in $\alpha$ and $C$. Such a sample-dependent occurrence of these magnetic transitions has already been found by different reports, including only one transition at $T_N \sim 13$ K [16], one transition at $T_N = 7$ K [14], three transitions at $T_{N1} \sim 14$ K, $T_{N2} \sim 10$ K and $T_{N3} \sim 8$ K[22], four transitions at $T_{N1} \sim 14$ K, $T_{N2} \sim 12$ K, $T_{N3} \sim 10$ K, $T_{N4} \sim 7.5$ K [17]. It has been suggested that different stacking sequences are responsible for this inconsistency, in which the transition near 14 K occurs in a $ABAB$ stacking of the hexagonal layers whereas the $ABC$ stacking produces the transition around 7 K [14]. This kind of stacking faults is inevitable due to the hysteric structural transition and consequently partial transformation of the sample structure as discussed above. We note that the structural transformation appears more complete in Sample 2 due to the larger $c$-axis length change at the transition [see Fig. 1(a) inset]. The magnetic transition at $T_{N3}$ in Sample 2 is also much stronger, and, therefore, we attribute the transition at $T_{N3}$ to the three-dimensional long-range magnetic order of the structurally transformed phase and the much weaker transition at $T_{N1}$ to the non-transforming fraction. We argue that the observed multiple magnetic transitions are not successive transitions in a well-defined single structural phase, but rather are likely due to the presence of several structural phases with different stacking sequences.

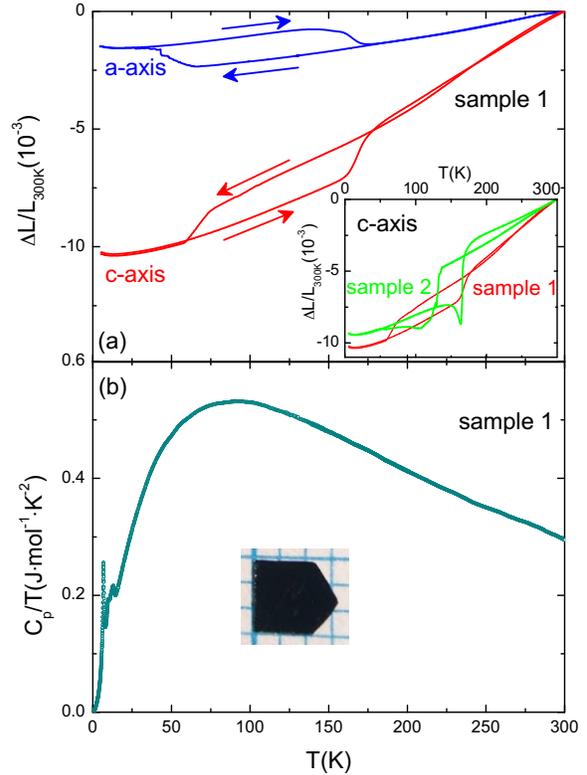

FIG. 1. (a) Temperature dependence of the thermal-expansion of $\alpha$-RuCl$_3$ along $a$ (blue curves) and $c$ (red curves) axes. A structural transition with a huge hysteresis is observed in both directions (arrow down: cooling, arrow up: heating). Inset: The hysteresis of the structural transition of a second sample (sample 2) is considerably smaller. (b) Specific heat versus temperature (upon cooling) clearly showing anomalies at the magnetic transition, however no clear anomalies at the structural transition. A photograph of Sample 1 is shown in the inset.

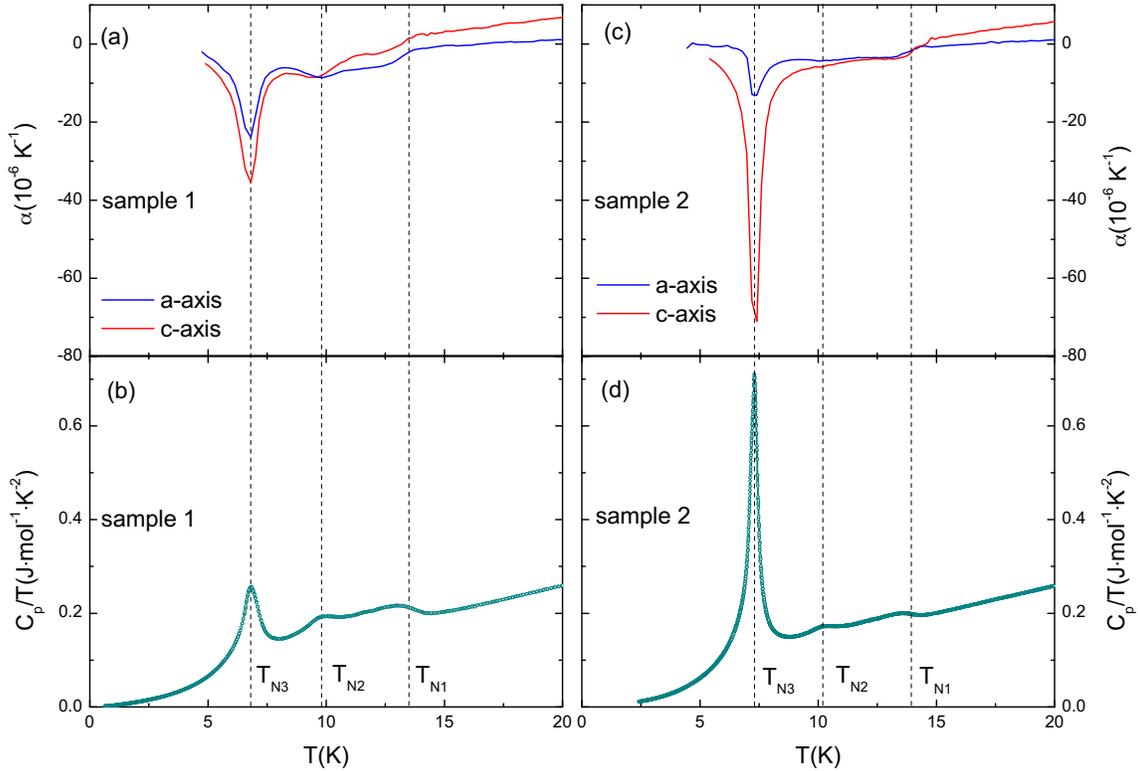

FIG. 2. Linear thermal-expansion coefficient $\alpha_i = \frac{1}{L_i}\frac{dL_i}{dT}$ and specific heat of two $\alpha$-RuCl$_3$ crystals at low temperatures. Temperature dependence of (a) linear thermal-expansion coefficient $\alpha$ and (b) specific heat $C/T$ of Sample 1. (c)(d) the same plot as (a)(b) for Sample 2. Vertical *dash lines* mark the magnetic transitions.

In the following we compute the pressure dependences of the magnetic transitions and compare our results to the non-trivial phase digram observed in hydrostatic pressure experiments [19, 25].

### C. Pressure dependence of the magnetic transitions

In this section we compute the pressure dependence of the magnetic transitions using the thermodynamic relation between the thermal-expansion and specific heat data shown in Fig. 2. For a first-order phase transition, the uniaxial pressure derivative of the phase transition temperature $T_c$ can be obtained via the Clausius-Clapeyron relation,

$$\frac{dT_c}{dp_i} = V_m \frac{\Delta L_i/L_i}{\Delta S}, \qquad (1)$$

where $\Delta L_i$ and $\Delta S$ are the discontinuities in the sample length and entropy at the phase transition, respectively, and $V_m = 53.32$ $cm^3/mol$ is the molar volume of $\alpha$-RuCl$_3$. At a second oder phase transition, the pressure dependence of the transition temperature is given by the Ehrenfest relation,

$$\frac{dT_c}{dp_i} = V_m \frac{\Delta \alpha_i}{\Delta C_p/T_c}, \quad (2)$$

in which $\Delta \alpha_i$ and $\Delta C_p$ are the jumps at $T_c$ in the linear thermal-expansion coefficient and specific heat, respectively. Interestingly, the transition at $T_{N1}$ exhibits mean-field type second- order phase transition behavior, i.e. a simple jump in $C$ and $\alpha$, implying long-range magnetic interactions. In contrast, the transitions at $T_{N2}$ and $T_{N3}$ exhibit the more typical behavior of magnetic systems with short-range interactions, i.e. continuous transitions with strong fluctuations in $C$ and $\alpha$ (see Fig. 2), for which the Pippard relation [28], in which one simply scales the respective anomalies in thermal expansion and heat capacity, should be applied. The absence of magnetic fluctuations above the transition at $T_{N1}$ in the linear thermal expansion and specific heat may be a manifestation of the Kitaev bond-directional interactions which suppresses the Heisenberg interactions in the $ABAB$ stacking type. For the $ABC$ staking, the zig-zag AF magnetic ground state established by sizable anisotropic Heisenberg interactions[14]. Hence, the structural details are essential to the ground state of $\alpha$-RuCl$_3$ and to eventually realize the quantum spin liquid state.

Table I. Pressure derivative of the three magnetic transitions in $\alpha$-RuCl$_3$. The effect at $T_{N2}$ of Sample 2 is very subtle in the thermal-expansion data, from which the pressure derivative can not be calculated reliably.

| $dT_N/dp_i$[K GPa$^{-1}$] | $dT_{N1}/dp_i$ | | | $dT_{N2}/dp_i$ | | | $dT_{N3}/dp_i$ | | |
|---|---|---|---|---|---|---|---|---|---|
| $i$ | $a$ | $c$ | $h$ | $a$ | $c$ | $h$ | $a$ | $c$ | $h$ |
| Sample 1 | -5.2(3) | -5.1(2) | -15.5(5) | -4.2(3) | -6.1(3) | -14.5(6) | -6.6(2) | -10.3(2) | -23.5(4) |
| Sample 2 | -3.6(3) | -6.2(3) | -13.4(6) | NA | NA | NA | -3.5(2) | -14.5(2) | -21.5(4) |

The uniaxial and hydrostatic pressure derivative of all three magnetic transitions, calculated using these thermodynamic relations using the thermal-expansion and specific heat data presented in Fig. 2, are presented in Table I. Since the anomalies in the thermal expansion are all of opposite sign as the heat capacity anomalies, the uniaxial pressure effects for all three transitions are negative. The hydrostatic pressure derivative is obtained by simply summing up the uniaxial components, e.g. $dT_N/dp_h = 2 \times dT_N/dp_a + dT_N/dp_c$, and is also negative for all transitions. We note that the transition at $T_{N2}$ in Sample 2 is very weakly pronounced, and we were thus not able to obtain reliable pressure derivatives.

### D. Phase diagram

The combined temperature-pressure phase diagram of the magnetic transitions occurring in the different polymorphs can be estimated by linearly extrapolating the initial pressure derivative of these transitions calculated in Section C and is displayed in Figure 3, together with direct hydrostatic pressure measurements [19, 25]. From our thermodynamic analysis, we predict that all transitions are suppressed strongly by hydrostatic pressure at roughly the same rate. This suggest that the microscopic physics of all three transitions, i.e. the exchange interaction, is quite similar for all three transitions. On the other hand, our results are at odds with the direct pressure measurements, which find essentially zero $dT_{N3}/dp$ for small pressures within the uncertainty of the data[19, 25]. In the heat capacity measurements performed under pressure by Wang et al. [19](red squares in Fig. 3), the magnetic transition at $T_{N3}$ is nearly constant up to $\sim$0.7 GPa, above which the transition suddenly disappears. In contrast, Cui et al. using NMR technique under pressure [25] (orange circles in Fig. 3) find that $T_{N3}$ first decreases slightly from 8 K down to 6 K at a pressure of 0.3 - 0.45 GPa and then increases up to $\sim$14 K at $\sim$1.1 GPa. At the same time, the volume fraction of the AFM phase decreases to zero above 1 GPa. In addition these authors observe a strongly pressure-dependent anomaly in the magnetization at higher temperatures (100 K - 250 K)[25]. In the following we discuss these very conflicting results in light of our present data.

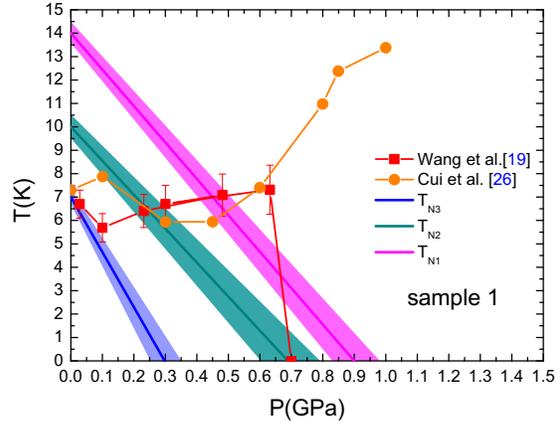

FIG. 3. Temperature-pressure phase diagram of the magnetic transitions in $\alpha$-RuCl$_3$. Lines: the pressure dependence of the three magnetic transition temperatures calculated from $dT_N/dp$ data of Sample 1 shown in Table I. The width of the lines represents the uncertainness in the calculations. Red squares [19] and orange circles [25] are taken from direct hydrostatic experiments.

We first note that the structural transition in our thermal-expansion data is very pronounced [see Fig. 1], whereas it is invisible in our heat capacity data, in agreement with previous heat capacity data [17]. This implies that the pressure derivatives of the structural transition are enormous according to Eq. 1. The implication of this result is that, in a real pressure experiment one probes not only the pressure derivative of the magnetic transitions, since pressure will dramatically change the state above $T_N$ out of which magnetism emerges, as demonstrated in Ref. [25]. On the other hand, the pressure derivatives obtained using our thermal expansion data represent the 'true' pressure derivatives of the magnetic transitions, since we do not apply any pressure and thus do not change the state out of which magnetic order emerges. Our results suggest that it should be possible to suppress the long-range magnetic order of the transition at $T_{N3}$ by a very moderate pressure of 0.3 GPa, if one can prevent a change of the crystal due to the high temperature structural transition. One possibility of realizing this may be to apply the pressure at very low temperatures, since the proposed stacking rearrangement occurring at the structural transition probably need some thermal activation in order to 'jump' over the energy barrier associated with this rearrangement.

## IV. CONCLUSIONS

Using thermal expansion and heat capacity measurements, we have shown that $\alpha$-RuCl$_3$ undergoes a strongly hysteretic structural transition, which is most likely responsible for the different reported magnetic transition temperatures in this system due to varying volume fractions of the low- and high-temperature structural phases. The structural transition is most likely related to a change of stacking sequence, which would naturally explain the large hysteretic behavior. We have argued that the magnetic transition at 7 K is associated with the structurally transformed phase at the high-temperature structural transition and thus represents the ground state of $\alpha$-RuCl$_3$. The transitions occurring at higher temperature most likely result from the non-transforming fraction. Using our data and thermodynamic relations, we have derived the uniaxial and hydrostatic pressure derivatives of the three magnetic transitions, associated with the different polymorphs. Our results suggest that long-range magnetic order should be totally suppressed by very moderate pressures of 0.3 GPa to 0.9 GPa. Our results are however at odds with real pressure measurements, which we attribute to the fact that in real pressure experiments, pressure strongly changes the state out of which magnetism emerges due to the large pressure dependence of the structural transition. In contrast, in our experiments, we probe the pressure effect upon magnetic order without applying any pressure and thus probe the 'true' pressure dependence of the system without changing the stacking. Our findings suggest that $\alpha$-RuCl$_3$ might still be an ideal playground to realize Kitaev physics using a moderate external pressure. This would however work only if one can prevent the pressure-induced stacking rearrangement, which may be possible if the pressure is applied at very low T to the presumably relatively large energy barrier involved in such a transition.